\documentclass{article}

\makeatletter
\def\and{%
  \end{tabular}%
  \hskip 0.5em \@plus.17fil\relax
  \begin{tabular}[t]{c}}
\makeatother

\usepackage[utf8]{inputenc}
\usepackage{amsmath, graphicx, apacite, xcolor}
\usepackage{indentfirst}
\usepackage{setspace}
\graphicspath{ {./images/} }
\usepackage{natbib}
\usepackage[a4paper,bindingoffset=0.2in,%
            left=1in,right=1in,top=1in,bottom=1in,%
            footskip=.25in]{geometry}
\newgeometry{left=1in,right=1in}

\begin{document}
\begin{titlepage}
    \title{The Hot Hand and Its Effect on the NBA}
    \author{
        McNair, Brian\\
        bmcnair@umich.edu
        \and 
        Margolin, Eric\\
        emargo@umich.edu
        \and
        Law, Michael \\
        mmylaw@umich.edu
        \and
        Ritov, Ya\hspace{-.1em}'\hspace{-.1em}acov \\
        yritov@umich.edu
        
    }
    \maketitle
    \begin{center}
    Department of Statistics\\
       University of Michigan\\
       Ann Arbor MI, USA
    \end{center}
   \begin{abstract}
       This paper aims to revisit and expand upon previous work on the ``hot hand" phenomenon in basketball, specifically in the NBA. Using larger, modern data sets, we test streakiness of shooting patterns and the presence of hot hand behavior in free throw shooting, while going further by examining league-wide hot hand trends and the changes in individual player behavior. Additionally, we perform simulations in order to assess their power. While we find no evidence of the hot hand in game-play and only weak evidence in free throw trials, we find that some NBA players exhibit behavioral changes based on the outcome of their previous shot. 
   \end{abstract}
\end{titlepage}
\doublespacing
\section{Introduction} \label{intro}
The “hot hand” phenomenon is at the center of a decades-long debate between basketball fans and social scientists. Many players, fans, coaches, and executives believe that a player who is ``hot" --- that is, they have made multiple consecutive successful shots --- is more likely to make their next shot attempt. As an example, on January 21st, 2019, Golden State Warriors guard Klay Thompson hit all 10 of his three-point attempts in a game against the Los Angeles Lakers, tying an NBA record for the most three pointers in a game without a miss. When asked about the feat afterwards in a post-game interview, Warriors head coach Steve Kerr said, ``Klay does that five or six times a season. You guys have seen it. He just got red hot, white hot" \citep{noauthor_klay_2019}.

In \citet*{gilovich_hot_1985} (henceforth referred to as GVT), the first academic research to test the widely accepted notion of the hot hand, the authors found no evidence supporting the phenomenon in basketball. Through the examination of NBA field goal data, NBA free throw data, and controlled shooting experiments with college players, the authors concluded that streaky shooting behavior, or  ``hotness", was nothing more than a belief in the fallacious law of small numbers \citep{tversky_belief_1971}, and that there lacked evidence to support the idea that players were truly more likely to make a shot after a make than after a miss. These findings led GVT to label the phenomenon a ``widely shared cognitive illusion."

Their ideas did not go published without criticism. \cite{larkey_its_1989} compiled their own data set and sought to undermine GVT's conclusions by analyzing streaks of high shooting percentages (instead of solely shooting streaks) and deriving their own statistics to conclude that total disregard of the hot hand in basketball was unreasonable. \cite{tversky_hot2_1989} responded by suggesting potential flaws in the \cite{larkey_its_1989} methods and logic, asserting that the paper's authors were misunderstanding the scope of their argument. Even today, as evident in Kerr's quote about Thompson, there is still a divide between those who accept the hot hand hypothesis, and those who side with GVT in denying it. Very recently, \cite{miller_surprised_2018} showed that the statistic used to estimate hot hand behavior (the empirical conditional probability of making a shot given the previous $k$ outcomes) is biased, and correcting for this bias in controlled shooting experiments, as they did with GVT's data, suggests plausibility in the hot hand after all. \cite{miller_is_2015} also showed support for the hot hand in three-point competitions, which is often thought of as a nice balance between a controlled shooting experiment and a competitive environment. The free throw line, a similar setting, might show some signs of hot hand shooting (\cite{arkes2010}, \cite{lantis_hot_2019}). 

The original debate, however, is rooted in whether shooting patterns observed in a live game can accurately be labeled as hot shooting behavior, which is difficult to detect. Multiple studies have tried to account for the external factors that get in the way of observing hot hand behavior, such as increased defensive pressure and varying shot difficulty, but there is still no  consensus as to whether or not the phenomenon witnessed on the court can truly be described as hot shooting (\cite{aharoni_hot_2012}, \cite{csapo_effect_2015}, \cite{lantis_hot_2019}, \cite{miller_cold_2014},  \cite{rao_experts_2009}). Consistent with surveys conducted in GVT, various papers have shown that shot difficulty tends to change in response to detecting hot shooting behavior through increased defensive pressure or different shot selection, which likely accounts for the difficulty in confirming or denying the presence of hot shooting in game situations (\cite{aharoni_hot_2012}, \cite{bocskocsky_heat_2014}, \cite{csapo_effect_2015}, \cite{rao_experts_2009}) \cite{aharoni_hot_2012} and \cite{csapo_effect_2015} argue that since shot difficulty increases during hot streaks while shooting percentage does not significantly change, the hot hand does exist in game settings, but this conclusion is inconsistent with the idea that a player exhibiting hot hand shooting is more likely to make their next shot, the central premise behind GVT's original paper. However, as it is difficult to detect the hot hand using massive data sets and precise statistical methods, one should question the ability of any observer of a basketball game to detect a phenomena.

Our paper aims to revisit, expand upon, and analyze the methods put forward in  previous papers, namely GVT. Using  modern datasets, we attempt to replicate the results of prior studies, as well as further expand upon analyses of changes in players behavior of the attacker as well as the defenders in response to perceived hot hand behavior. Additionally, due to the difficulties in detecting the hot hand in game action, we also run simulations of some of our methods to determine how frequently we would expect to detect the hot hand in order to assess the statistical power of the procedure.

In the end, there is no evidence of a hot hand phenomenon in the field of play, and only weak evidence of a hot hand in free throw shooting. However, we detected changes in player behavior in response to their previous shot outcomes, leading us to question GVT's initial shot independence assumption. This combination leads us to conclude that GVT's outright rejection of the hot hand is based on limited information and can thus only reject the hot hand in an equally limited capacity. 
The second section of our paper provides an overview of our main data sources, and the third section contains our analysis, followed by our conclusions, tables, and figures.

\section{Data} \label{data}
Two datasets were utilized in this study: a dataset of shots from the 2014--15 NBA season from the NBA API (which can be found publicly at: https://www.kaggle.com/dansbecker/nba-shot-logs), as well as a dataset of free throws from the 2018--19 NBA season, scraped from Basketball-Reference.com. The NBA API dataset (as it will be referred to henceforth) contains 205,539 shots with many features, including shooter, shot distance, number of dribbles before shot attempt, distance to nearest defender, and type of shot, among many other variables. For this paper, we make use of the aforementioned variables.

The free throw data from Basketball-Reference consists of 59,345  free throw attempts from 582 players, with free throw trips from flagrant (or technical) fouls filtered out. Each row consists of a game identification code, the free throw shooter, the result of the attempt, the number of attempts the shooter had in their trip, and the score of the game. This dataset was filtered to only include trips to the free throw line with at least two attempts, which lowered the number of free throws considered to 52,979. 

\section{Analysis} \label{analysis}
\subsection{Analysis of Runs} \label{analysisofruns}
The core idea to the hot hand phenomenon is the belief that a player has periods during games in which he is shooting more accurately, and is thus more likely to hit their next shot after hitting multiple shots in a row. If this idea were true, then each shot would not be independent with one another. GVT tested this hypothesis with a Wald-Wolfowitz runs test \citep{wald_exact_1943}. Using the NBA API dataset, a Wald-Wolfowitz test was conducted on the shooting sequences of 443 players to see if streaky behavior was present. Streaks were limited to within games, thus the streak at the end of one game was not concatenated with the streak at the beginning of the next game. 

 With this dataset, 25 out of 443 players (5.6\%) had significant p-values at the 5\% level, which is within the range of expected significant p-values of (3.2\%, 7.1\%) under the null hypothesis. These results do not lend themselves to the existence of an observed hot hand phenomenon, which is consistent with the findings in GVT. However, the Wald-Wolfowitz tests look for the hot hand in each individual player. Our next section examines global hot hand behavior, in order to present a clearer league-wide picture.

\subsection{Global test statistic} \label{globalteststatistic}

Before continuing any further, it is important for us to provide a slightly more elaborate quantification of the ``hot hand." Let $p_{M_i}$ represent the probability of a player $i$ making a shot after a miss, and $p_{H_i}$ represent the probability of that same player making a shot after a make. Then, the effect size of the hot hand for player $i$ can be estimated by $\hat{p}_{H_{i}} - \hat{p}_{M_{i}}$. Further, if the hot hand phenomenon was real, it would be expected that $p_{H_{i}} - p_{M_{i}} > 0$, as this would signal a player is more likely to make a shot after a make than after a miss. Each player's shots in the NBA API dataset were separated into disjoint pairs of consecutive shots to avoid the criteria of MLS within each game and categorized as one of the following: a “hit-hit” pair when the player made both shots, a “hit-miss” when they made the first shot and missed the second, “miss-miss,” and “miss-hit” for the corresponding outcomes. Each player's  $\hat{p}_{H_i}$, the proportion of made shots after the first shot was made, and $\hat{p}_{M_i}$, the proportion of made shots after the first shot was missed, were calculated. Under the null hypothesis, this estimator has a variance of $1/n_i$, where $n_i$ is the total number of disjoint pairs for player $i$. 
 
Then, to globally test if a player's shooting percentage was independent of their preceding shot attempt, the following test statistic was utilized, aggregating the data for each player $i$ ($N = 443$ players):
\begin{align}
    T = \frac{1}{\sqrt{\sum_{i = 1}^{N}{n_i}}} \sum_{i = 1}^{N} n_i (\hat{p}_{H_{i}} - \hat{p}_{M_i}),
\end{align}
which has an asynptotic, non-standard normal distribution. 

A global test statistic of -4.4396 was computed from the disjoint pairs. The mean $\hat{p}_{H_{i}}$ value was 0.4477, while the mean $\hat{p}_{M_{i}}$ value was 0.4554. As we were testing the one-sided hypothesis that $p_{H_{i}} - p_{M_i} > 0$, this negative test statistic is highly insignificant. While hot hand characteristics may occasionally be found in some players on an individual level, these results do not indicate a trend of a general hot hand throughout the entire NBA. There are other parts of the game, however, where the hot hand could exist as a global phenomenon.

\subsection{Free Throws} \label{freethrows}

To remove the possible influences that defensive behavior and shot selection might have on the shot outcome, GVT examined the serial correlation of free throw data from nine players on the 1981--82 Boston Celtics. They aimed to check if the hot hand effect was present in a fixed competitive environment. By examining each player's correlation between the outcomes of their first free throw attempts and the outcomes of their second attempts, GVT concluded there was no evidence of hot hand behavior within free throw shooting. Using our recent free throw data, the same procedure was conducted. The correlations between the outcome of the first shot versus the outcome of the second shot were recorded for 480 players who had more than one trip to the free throw line, and each value was tested to see if it was significantly greater than zero. Table \ref{table:X} shows the test results for players with the highest amount of free throw trips among significant correlation values.

Overall, a total of 37 out of 480 players had serial correlation values that were significantly greater than zero at a 5\% significance level, and therefore exhibited some hot hand behavior, which is slightly above the range of expected significant p-values of (14.64, 33.36) under the null, suggesting weak support of the hot hand effect. These results are slightly different than GVT, but are similar to findings from  \cite{arkes2010} and \cite{lantis_hot_2019}. It is worth noting that there may be debate as to whether free throw shooting (as well as other controlled shooting environments) are truly relevant to the hot hand discussion, since free throws are exact mechanical repetitions, and observers normally do not acknowledge or care whether a free throw shooter is ``hot," with the exception of incredibly long streaks. Nonetheless, hot hand shooting is only weakly found in free throw shooting, and these results are not very relevant in the more popular conversation of whether the hot hand exists in the field of play.

\subsection{Player Behavior in Relation to the Hot Hand} \label{playerbehavior}

Many defenders of the hot hand phenomenon point to differences ``in both teams' behavior after the detection of the hot hand" as a challenge in observing the hot hand's  existence (\cite{aharoni_hot_2012}, \cite{csapo_effect_2015}). They argue that defensive players focus more attention on hot players while offensive players focus more on getting hot players the ball. Both of these changes directly impact the probability a hot player makes their next shot and thus need to be considered in any analysis of the hot hand phenomenon. Using the NBA API data set, we attempt to quantify the aforementioned behavioral changes.

\subsubsection{Shot Distance} \label{shotdistance}
The phrase ``heat check” is thrown around by NBA analysts very frequently. It describes the instance where a hot player takes a ``low percentage" shot (most frequently a long three-pointer), which for most players, ends their streak of makes. More generally, if shooters take  shots from further away after a make than after a miss (a shot which they are more likely to miss), it could partially explain why the hot hand phenomenon would not be seen in tests that did not factor shot difficulty into account.   

For every player in our data set, we calculated the mean distance of their shot after  a make and the mean distance of their shot after a miss. In order to eliminate the ``carry over effect" from game to game we re-coded each player's first shot of the game as an unknown value, allowing us to better quantify the hot hand. Contrary to widespread intuition, our analysis showed that 170 players took longer shots after a make and 171 players took longer shots after a miss. Among the 341 players analyzed, 19 ($5.5\%$ CI: $2.83\%, 8.17\%$) had a significant difference between the two means, with six ($1.7\%$ CI: $0.33\%, 3.07\%$) taking shots from further after a make and 13 ($3.8\%$ CI: $1.56\%, 6.04\% $) taking shots from further after a miss. Those players with significant differences are included in Table \ref{table:X3}.  Using the Benjamini-Hochberg Procedure \citep{benjamini_controlling_1995} to control False Discovery Rate (FDR) with a significance level of 0.05, we cannot confidently declare any discoveries for any of the p-values.

These results go against the common hot hand intuition that players tend to be eager to attempt longer shots when they are on a hot streak. They instead show an insignificant mixed bag, with some players even attempting longer shots after miss. It is important to note that certain positions have different shooting behaviors, and it might be insightful to look into how these results vary by position. Again, even if the phenomenon exists, it is impossible to detect without sophisticated tools. In general, however, the outcome of previous shots doesn't seem to significantly alter player shot attempts throughout the league as a whole. 

\subsubsection{Shot Frequency} \label{shotfrequency}
Another way to quantify an increase in ``risky behavior" due to increased confidence is the frequency with which a player shoots. One would expect a player who just made a shot would take their next shot sooner than one who just missed. The time between shot attempts (measured in ``game time``) for each player was used to examine whether players tended to take shots more frequently after a make than after a miss.

In order to quantify this phenomenon, we used a similar methodology as in \ref{shotdistance}, calculating  the mean time before attempting another shot after a make and after a miss. Among the 341 players analyzed, 148 ($43\%$) players took shots later after a make, and 193 ($57\%$) players took shots earlier after a make. 25  ($7.3 \%$ CI: $4.54\%, 10.06\%$) had a significant difference between the two means, with nine ($2.6\%$ CI: $0.91\%, 4.29\%$) taking shots sooner after a make and 16 ($4.6 \%$ CI: $2.38\%, 6.82\%$) taking shots sooner after a miss. Those players with significant differences are included in Table \ref{table:X4}. Using the Benjamini-Hochberg Procedure to control FDR, for an $\alpha$ of 0.05, we can confidently reject only one of the null hypotheses tested.

These results, combined with the findings of \ref{shotdistance} seem to counter the ``heat-check" concept associated with the hot hand phenomenon. Players seem to not be influenced by the success of their previous shots in any systematic way. If anything, specific players seem to be more judicious after a make than a miss as seen in the generally closer shots taken after longer intervals.

\subsubsection{Dribbles Before A Shot} \label{dribbles}
Ball movement is often regarded as a crucial skill of a winning team \citep{d2015move}. While the datasets used in this study did not include data on the number of passes a player makes, we can use their number of dribbles as a proxy to arrive at a similar measure of ball movement within an NBA offense. Using the NBA API data, we use a similar methodology as \ref{shotdistance} and \ref{shotfrequency}, calculating the mean number of dribbles a player takes after a made shot versus a missed shot. 
Out of 281 players with a significant number of shots, 102 ($36.3\%$) players take fewer dribbles after a make and 179 ($63.7\%$) take fewer dribbles after a miss. Forty ($14.2\%$ CI: $10.12\%, 18.28\%$) of these players had significant differences between their average number of dribbles, with six  ($2.1\%$ CI: $0.42\%, 3.78\%$) taking more dribbles after a miss and 34 ($12.1\%$ CI: $10.12\%, 18.28\%$) taking more dribbles after a make. Those players with significant differences are included in Table \ref{table:X5}. Using the Benjamini-Hochberg Procedure to control FDR, for an $\alpha$ of 0.05, we can confidently reject one of the null hypotheses tested. 

From the discussions in \ref{shotdistance} and \ref{shotfrequency} we can see that, generally, NBA players do not tend to take ``riskier" shots after a make. This, however, does not mean that there aren't changes in an offensive player's behavior based on the outcome of a previous shot. Our analysis of dribbles, shows that a minority of offensive players do behave differently, taking more dribbles, on average, after a make than after a miss. 

\subsubsection{Defender Distance} \label{defenderdistance}
Thus far we have solely focused on changes in offensive behavior as a result of a made shot. Yet, the five  players on the other team have an obvious impact on the game and should thus be considered when analyzing player behavior in regards to the hot hand. In essence, we hope to answer the question ``Is a player guarded more closely after a make than a miss?"

Using the same 281 players from sections \ref{shotdistance}--\ref{dribbles}, we calculated the mean distance from the closest defender for each shot taken after a miss and after a make. 175 ($62.3\%$) players had defenders closer after a make and 106 ($37.7\%$) had defenders closer after a miss. Of these players, 34 ($12.1\%$  CI: $8.37\%, 16.03\%$) had significant differences in their mean defender distance. Seven ($2.5\%$ CI: $0.67\%, 4.33\%$) of these significant players had defenders closer after a miss and 27 ($9.6\%$  CI: $6.16\%, 13.04\%$) had defenders closer after a make. Those players with significant differences are included in Table \ref{table:X6}.

Using the Benjamini-Hochberg Procedure \citep{benjamini_controlling_1995} to control FDR, for an alpha of 0.05, we can confidently reject six of the null hypotheses tested. These results suggest that globally, defenders do not appear to play significantly ``tighter" or ``looser" defense on an opponent depending on the result of the previous shot.

\subsubsection{Discussion on Player Behavior and the Hot Hand} \label{behaviordiscussion}
This section is intended to investigate the claim that basketball players change their behavior based on the result of their previous shot. From our analysis we see that, despite a few exceptions, NBA players do not tend to change their behavior based on the previous performance. This macro-observation debunks a crucial tenant of the pro-hot-hand argument, increasing confidence in GVT's initial claim that the hot hand is mainly an observed psychological phenomenon.  

While no players have significant differences in all three of the offensive metrics tested, three players (Dwayne Wade, Jarret Jack, and John Wall) had significant differences in both the number of dribbles taken before a shot and shot frequency after a miss versus after a make, and two players (James Jones and Al-Farouq Aminu) have significant differences in both their shot frequency and shot distance after a miss versus a make. Thus, there appears to be some support for further examination of the idea that a few offensive players do change their behavior significantly in response to perceiving the hot hand, which is consistent with the findings in GVT's survey of basketball players, but not factored in to their in-game testing methodology.

\subsection{Impact of Game Breaks on Hot Hand} \label{gamebreaks}
If the hot hand does exist in the field of play, then it is in the best interest of the opposition to devise strategies aimed to ``cool" off a player perceived to be hot. Coaches and players might like to know how quickly a player can slip out of hot behavior, or how easily a player can ``cool" down when starting a hot streak. Aside from readjusting defensive pressure, a common strategy utilized by coaches trying to stop players from continuing their apparent hot streaks is to influence the flow of the game. This idea can be focused in offensive play, as some teams believe that changing the speed of their offense (and how deep into a possession they attempt a shot) will help them control the tempo of the game, which is believed to influence the performance of the opposition. Whether or not there is viability to this belief, a more direct way of influencing game flow is through play stoppage, via a timeout or end of a period. One form of this strategy, often referred to as ``icing a player", occurs when a time-out is called in-between an opposing player's free throw attempts. Icing is practiced in similar ways throughout different sports, such as with field goal kicking in American football. The idea is that by increasing the time between attempts, it not only gives the shooter more time to think about their shot and get ``psyched out", but it removes them from their supposed rhythm that underlies their shooting performance. Regardless of whether or not the hot hand does exist, it is still useful for a team to know how they can instigate cold shooting behavior in an opponent, especially in critical game situations where the outcome of a game rests on one specific shot. Thus, there is value in studying the impact of these strategies.

To test how breaks in game action impact a player's shooting performance, field goal percentages before and after halftime were analyzed. Although these situations do not mirror free throw attempts and do not illustrate the effect of ``icing a shooter", they do offer a glimpse as to how a substantial break from action affects shooting. In the NBA, halftime is fifteen minutes long, and is the longest pause from play throughout the entire game. Filtering the NBA API dataset to just performances by players who attempted at least three shots in the second and third quarters, player shooting probabilities between the second and third quarters were analyzed. The correlation results between field goal percentage just before the half and field goal percentages just after the half with different subsets of data are found in Table \ref{table:X7}. There does not appear to be strong evidence that shooting performance changes significantly after the halftime break. There was one significant p-value at the 5\% level with the subset of shots consisting of the last five attempts in quarter two and the first five shots in quarter three, but the global field goal percentages in those two subsets were nearly identical, but every other subset had a highly insignificant p-value. There does not appear to be strong evidence in favor of the halftime break instigating cold shooting behavior in players. Further research into the effects of timeouts on shooting behavior, especially in icing situations, would be beneficial to expand upon this analysis.
 
\subsection{Simulations} \label{sims}
In order to evaluate power, simulations were performed to assess when hot hand behavior would be detected with the procedures used in this study. Using the NBA API dataset, for all $k_{it}$ shots attempted by player $i$ in quarter $t$ (if player $i$ attempted a shot that quarter), $k_{it}$ Bernoulli random variables were generated, where the probability of player $i$ making a shot in quarter $t$ was generated by the following:
\begin{align}
    \begin{cases}
    p_{i} / (1 - \delta) & \text{with probability } (1-\delta)/4, \\
    p_{i} & \text{with probability } 0.5, \\
    p_{i} / (1 + \delta) & \text{with probability }  (1+\delta)/4,
    \end{cases}
\end{align}
where $p_{i}$ is player $i$'s seasonal field goal percentage. A new probability was generated for every quarter player $i$ recorded a shot. The value of $\delta$ was changed for each set of simulations, with the objective of identifying which value of $\delta$ was necessary to start detecting hot hand behavior, where a $\delta$ value of 0 theoretically emulates the actual data. As $\delta$ increases, we expected it to have an increasing effect on our simulated results, as the gap between the higher shooting probability and the lower shooting probability widen. For each value of $\delta$, which ranged from values of zero to 0.6 (inclusive) with values differing by 0.03, 10 simulations were conducted, resulting in a total of 210 simulations generated. 

\subsubsection{Analysis of Runs} \label{analysisofrunssims}
The results of running the Wald-Wolfowitz procedure on each player with Benjamini-Hochberg on the simulated data can be found in Figure \ref{fig:my_label}. With the discoveries being declared at a 5\% significance level, there is a strong positive relationship between the number of declared discoveries and the associated $\delta$ value. 
A significant number of discoveries (more than 22 declared discoveries) is first found at $\delta = 0.39$, and  38 percent of the simulations found strong evidence of the hot hand using the Wald-Wolfowitz procedure.

\subsubsection{Global Test Statistic} \label{globaltestsim}
Figure \ref{fig:my_label2} shows the results of finding the global test statistic in the simulations. As the value of $\delta$ increases, the value of $T$ constantly increases, which would correspond to moving in the direction of more significance ($H_0: T = 0$ versus $H_1: T > 0$). The first simulation results with a strong indication of simulated hot hand behavior was found at a $\delta$ around $0.18$. Overall, both simulations suggest that a $\delta$ value between $0.18$ and $0.39$ was necessary to start finding strong evidence of the hot hand using our methods.

\section{Conclusion}
Overall, we did not find any strong evidence in defense of the hot hand phenomenon. While we found a slight association between a player’s first free throw attempt and their second free throw attempt, similar to findings in \cite{arkes2010} and \cite{lantis_hot_2019}, there was not much evidence in favor of a hot hand effect in the field of play, the setting in which the debate is primarily rooted. And while it is becoming more evident that controlled shooting environments, like free throw attempts and the NBA's three-point contest, do foster hot hand shooting (\cite{miller_is_2015},  \cite{miller_surprised_2018}), it is difficult to find the existence of hot shooting in real game scenarios. If the debate is about whether the shooting patterns observed on the court show traces of the hot hand, then our analysis would add on to existing studies in dissenting, though the more common question in modern statistical literature on the subject is whether there is hotness when controlling for shot difficulty and other external factors. Our study of game breaks suggests that pauses in game play may not have a significant effect on shooting percentage, though this finding does not necessarily apply to instances of ``icing" a shooter, such as in crucial free throw shooting situations.

Additionally, our analysis confirmed that there appears to be some changes in playing behavior as a result of a player’s previous shot, namely in the number of dribbles a player take before attempting another shot, though some players did appear to significantly change their general offensive behavior in response to their previous shot. This is consistent with findings in \cite{aharoni_hot_2012}, \cite{bocskocsky_heat_2014}, \cite{csapo_effect_2015}, and \cite{rao_experts_2009}, as well as survey results from GVT. Our simulations indicate that hot hand shooting is unlikely to be found with our methods, as adding a large effect size was necessary in order to detect hot shooting behavior in our simulations.

Three and a half decades after GVT's original paper, the hot hand debate is still ongoing. Initially, statisticians and psychologists rejected the phenomenon, while the basketball community continued to believe in it. However, recent literature in econometrics has since seen some changed academic perspectives, with some researchers arguing for the hot hand's existence in game settings, though this belief is not yet met with a consensus in the statistical community. With the many difficulties associated with controlling for shot difficulty, and the questionable relevance of these controls in answering the original question posed in GVT's paper, it is unclear if or when this debate will ultimately be settled.

\section{Acknowledgements}
This paper was supported in part by NSF grants DMS 1646108 and 1712962.  We would also like to offer special thanks to the Statistics Department at the University of Michigan for providing us the rare opportunity to work on a meaningful research project as undergraduates. 
Finally, we wish to thank our friends and families for their support and encouragement throughout this process.

\bibliography{refs}

@article{wald_exact_1943,
	title = {An exact test for randomness in the non-parametric case based on serial correlation},
	volume = {14},
	issn = {0003-4851},
	url = {http://projecteuclid.org/euclid.aoms/1177731358},
	doi = {10.1214/aoms/1177731358},
	language = {en},
	number = {4},
	urldate = {2020-08-05},
	journal = {The Annals of Mathematical Statistics},
	author = {Wald, A. and Wolfowitz, J.},
	month = dec,
	year = {1943},
	pages = {378--388}
}

@article{csapo_effect_2015,
	title = {The effect of perceived streakiness on the shot-taking behaviour of basketball players},
	volume = {15},
	issn = {1746-1391, 1536-7290},
	url = {http://www.tandfonline.com/doi/full/10.1080/17461391.2014.982205},
	doi = {10.1080/17461391.2014.982205},
	language = {en},
	number = {7},
	urldate = {2020-08-04},
	journal = {European Journal of Sport Science},
	author = {Csapo, Peter and Avugos, Simcha and Raab, Markus and Bar-Eli, Michael},
	month = oct,
	year = {2015},
	pages = {647--654}
}

@article{rao_experts_2009,
	title = {Experts’ {Perceptions} of {Autocorrelation}: {The} {Hot} {Hand} {Fallacy} {Among} {Professional} {Basketball} {Players}},
	journal = {Working Paper},
	author = {Rao, Justin},
	year = {2009}
}

@techreport{bocskocsky_heat_2014,
	address = {Rochester, NY},
	type = {{SSRN} {Scholarly} {Paper}},
	title = {Heat check: new evidence on the hot hand in basketball},
	shorttitle = {Heat check},
	url = {https://papers.ssrn.com/abstract=2481494},
	language = {en},
	number = {ID 2481494},
	urldate = {2020-08-04},
	institution = {Social Science Research Network},
	author = {Bocskocsky, Andrew and Ezekowitz, John and Stein, Carolyn},
	month = aug,
	year = {2014}
}

@article{benjamini_controlling_1995,
	title = {Controlling the false discovery rate: a practical and powerful approach to multiple testing},
	volume = {57},
	issn = {00359246},
	shorttitle = {Controlling the false discovery rate},
	url = {http://doi.wiley.com/10.1111/j.2517-6161.1995.tb02031.x},
	doi = {10.1111/j.2517-6161.1995.tb02031.x},
	language = {en},
	number = {1},
	urldate = {2020-07-24},
	journal = {Journal of the Royal Statistical Society: Series B (Methodological)},
	author = {Benjamini, Yoav and Hochberg, Yosef},
	month = jan,
	year = {1995},
	pages = {289--300}
}

@article{lantis_hot_2019,
	address = {Cambridge, MA},
	title = {Hot shots: an analysis of the ‘hot hand’ in nba field goal and free throw shooting},
	shorttitle = {Hot shots},
	url = {http://www.nber.org/papers/w26510.pdf},
	language = {en},
	number = {w26510},
	urldate = {2020-07-29},
	institution = {National Bureau of Economic Research},
	author = {Lantis, Robert and Nesson, Erik},
	month = nov,
	year = {2019},
	doi = {10.3386/w26510},
	pages = {w26510}
}

@article{larkey_its_1989,
	title = {It's okay to believe in the “hot hand”},
	volume = {2},
	issn = {0933-2480, 1867-2280},
	url = {http://www.tandfonline.com/doi/full/10.1080/09332480.1989.10554950},
	doi = {10.1080/09332480.1989.10554950},
	language = {en},
	number = {4},
	urldate = {2020-07-24},
	journal = {CHANCE},
	author = {Larkey, Patrick D. and Smith, Richard A. and Kadane, Joseph B.},
	month = sep,
	year = {1989},
	pages = {22--30}
}

@article{tversky_belief_1971,
	title = {Belief in the law of small numbers.},
	volume = {76},
	issn = {1939-1455, 0033-2909},
	url = {http://doi.apa.org/getdoi.cfm?doi=10.1037/h0031322},
	doi = {10.1037/h0031322},
	language = {en},
	number = {2},
	urldate = {2020-07-24},
	journal = {Psychological Bulletin},
	author = {Tversky, Amos and Kahneman, Daniel},
	year = {1971},
	pages = {105--110}
}

@article{tversky_hot2_1989,
	title = {The “hot hand”: statistical reality or cognitive illusion?},
	volume = {2},
	issn = {0933-2480, 1867-2280},
	shorttitle = {The “hot hand”},
	url = {http://www.tandfonline.com/doi/full/10.1080/09332480.1989.10554951},
	doi = {10.1080/09332480.1989.10554951},
	language = {en},
	number = {4},
	urldate = {2020-07-24},
	journal = {CHANCE},
	author = {Tversky, Amos and Gilovich, Thomas},
	month = sep,
	year = {1989},
	pages = {31--34}
}

@article{arkes2010,
  title = {Revisiting the Hot Hand Theory with Free Throw Data in a Multivariate Framework},
  author = {Arkes,, Jeremy},
  year = {2010},
  URL = {https://calhoun.nps.edu/handle/10945/43641},
  publisher = {Calhoun}
}

@misc{noauthor_klay_2019,
    author = "Nick Friedell",
	title = {Klay ties {NBA} mark with 10 straight 3-pointers},
	url = {https://www.espn.com/nba/story/_/id/25821022/klay-thompson-golden-state-warriors-ties-nba-record-10-straight-3-pointers},
	language = {en},
	urldate = {2020-07-24},
	journal = {ESPN.com},
	month = jan,
	year = {2019}
}

@article{aharoni_hot_2012,
	title = {Hot hands and equilibrium},
	volume = {44},
	issn = {0003-6846, 1466-4283},
	url = {http://www.tandfonline.com/doi/abs/10.1080/00036846.2011.564141},
	doi = {10.1080/00036846.2011.564141},
	language = {en},
	number = {18},
	urldate = {2020-07-24},
	journal = {Applied Economics},
	author = {Aharoni, Gil and Sarig, Oded H.},
	month = jun,
	year = {2012},
	pages = {2309--2320}
}

@article{gilovich_hot_1985,
	title = {The hot hand in basketball: {On} the misperception of random sequences},
	volume = {17},
	issn = {00100285},
	shorttitle = {The hot hand in basketball},
	url = {https://linkinghub.elsevier.com/retrieve/pii/0010028585900106},
	doi = {10.1016/0010-0285(85)90010-6},
	language = {en},
	number = {3},
	urldate = {2020-07-24},
	journal = {Cognitive Psychology},
	author = {Gilovich, Thomas and Vallone, Robert and Tversky, Amos},
	month = jul,
	year = {1985},
	pages = {295--314}
}

@article{miller_cold_2014,
	title = {A cold shower for the hot hand fallacy},
	issn = {1556-5068},
	url = {http://www.ssrn.com/abstract=2450479},
	doi = {10.2139/ssrn.2450479},
	language = {en},
	urldate = {2020-07-24},
	journal = {SSRN Electronic Journal},
	author = {Miller, Joshua Benjamin and Sanjurjo, Adam},
	year = {2014}
}

@article{miller_is_2015,
	title = {Is it a fallacy to believe in the hot hand in the nba three-point contest?},
	issn = {1556-5068},
	url = {http://www.ssrn.com/abstract=2611987},
	doi = {10.2139/ssrn.2611987},
	language = {en},
	urldate = {2020-07-24},
	journal = {SSRN Electronic Journal},
	author = {Miller, Joshua Benjamin and Sanjurjo, Adam},
	year = {2015}
}

@article{miller_surprised_2018,
	title = {Surprised by the hot hand fallacy? {A} truth in the law of small numbers},
	volume = {86},
	issn = {0012-9682},
	shorttitle = {Surprised by the hot hand fallacy?},
	url = {https://www.econometricsociety.org/doi/10.3982/ECTA14943},
	doi = {10.3982/ECTA14943},
	language = {en},
	number = {6},
	urldate = {2020-07-24},
	journal = {Econometrica},
	author = {Miller, Joshua B. and Sanjurjo, Adam},
	year = {2018},
	pages = {2019--2047}
}

@inproceedings{d2015move,
  title={Move or die: How Ball Movement Creates Open Shots in the NBA},
  author={D’Amour, Alexander and Cervone, Daniel and Bornn, Luke and Goldsberry, Kirk},
  booktitle={Sloan Sports Analytics Conference},
  year={2015}
}

\newpage
\section{Tables}

\begin{table}[h!]
\centering
\textbf{Sample of Players Exhibiting Hot Hand Behavior in Free Throws}\par\medskip
\begin{tabular}{rrrrr}
  \hline
 Player & $P(H_2|H_1)$ & $P(H_2|M_1)$ & $r$ & P-value \\ 
  \hline
  Kevin Durant & 0.86 (234) & 0.76 (37) & 0.10 & $4.7 \times 10^{-2}$ \\ 
Julius Randle & 0.82 (152) & 0.61 (59) & 0.22 & $5.3 \times 10^{-4}$ \\ 
  Andre Drummond & 0.73 (108) & 0.49 (89) & 0.24 & $2.9 \times 10^{-4}$ \\ 
   Montrezl Harrell & 0.76 (107) & 0.64 (75) & 0.14 & $3.2 \times 10^{-2}$\\ 
    LeBron James & 0.79 (111) & 0.57 (67) & 0.24 & $6.1 \times 10^{-4}$ \\ 
  Jimmy Butler & 0.86 (141) & 0.67 (21) & 0.17 & $5.3 \times 10^{-2}$ \\ 
   \hline
\end{tabular}
\caption{(\ref{freethrows}) The player's conditional probabilities of (H) hitting their second free throw attempt given making and missing their first attempts are listen in columns 2 and 3, respectively. The number of attempts are listed in parentheses. The serial correlation between the first and second attempts are listed in column four. The p-value of the hypothesis test $H_0: r = 0$ versus $H_1: r > 0$ is listed in column 5. The players listed had the five highest numbers of free throw trips (sorted in descending order) with significant serial correlations ($\alpha = 0.05$)}.
\label{table:X}
\end{table}

\begin{table}[h!]
\begin{center}
\textbf{Mean Shot Distance After a Make vs. After a Miss}\par\medskip
\begin{tabular}{rrrrr}
\hline
Player & Avg. Shot Distance After Make & Avg. Shot Distance After Miss & Z     & P-value  \\
\hline
Thabo Sefolosha   & 8.43(70)               & 13.09(67)              & -2.72 & 0.0065 \\
Mike Muscala      & 8.84(19)               & 16.08(26)              & -2.69 & 0.007  \\
Chris Johnson     & 11.40(20)               & 18.25(28)              & -2.59 & 0.0096 \\
Lance Thomas      & 11.37(108)              & 14.48(112)              & -2.50  & 0.0124 \\
Rajon Rondo       & 9.42(134)               & 11.86(155)              & -2.31 & 0.0209 \\
Rodney Hood       & 14.51              & 16.81              & -2.28 & 0.0228 \\
JaMychal Green    & 6.27(62)               & 9.86(59)               & -2.26 & 0.0238 \\
Dirk Nowitzki     & 15.67(177)              & 17.24(220)              & -2.18 & 0.0289 \\
Dewayne Dedmon    & 2.84(19)               & 6.00(21)                  & -2.15 & 0.0312 \\
Rudy Gay          & 9.82(190)               & 11.67(231)              & -2.13 & 0.0328 \\
Al Jefferson      & 7.54(61)               & 9.73(67)               & -2.09 & 0.0363 \\
Wesley Matthews   & 16.88(113)              & 19.03(175)              & -2.08 & 0.0372 \\
James Jones       & 18.96(28)              & 22.58(38)              & -2.06 & 0.039  \\
Langston Galloway & 16.61(87)              & 13.83(127)              & 1.99  & 0.0463 \\
Brian Roberts     & 18.30(20)               & 13.70(23)               & 2.02  & 0.0435 \\
Al-Farouq Aminu   & 15.43(129)             & 13.01(184)              & 2.08  & 0.0376 \\
Tayshaun Prince   & 15.53(45)             & 12.13(38)              & 2.10   & 0.0355 \\
Stanley Johnson   & 14.01(97)              & 11.01(168)              & 2.18  & 0.029  \\
Solomon Hill      & 13.32(19)              & 4.89(19)               & 2.78  & 0.0054 \\
\hline
\end{tabular}
\end{center}
\caption{(\ref{shotdistance}) The entire NBA API dataset was used in this analysis (Number of Players = 341). The second and third columns show the mean distance (in feet) of a player's next shot after a miss versus after a hit, with the number attempts in parenthesis. The fourth and fifth columns show Z-scores found comparing these two means and the associated p-values using the hypothesis test $H_0:  \mu_{hit} = \mu_{miss}$ vs. $H_1: \mu_{hit} \neq \mu_{miss}$. The table above displays those players with significant p-values ($\alpha = 0.05$).}
\label{table:X3}
\end{table}

\begin{table}[h!]
\centering
\textbf{Shot Frequency After a Make vs. After a Miss}\par\medskip
\begin{tabular}{rrrrr}
\hline
Player Name      & Avg. Time After Make & Avg. Time  After Miss & Z     & P-value  \\
\hline
Ronnie Price     & 175.13(31)                     & 486.82(50)                     & -3.81 & 0.0001   \\
Dwyane Wade      & 141.84 (209)                    & 192.50(251)                      & -2.76 & 0.0058 \\
Larry Nance Jr.  & 246.09(67)                     & 462.45(53)                     & -2.72 & 0.0065 \\
Trevor Ariza     & 221.52(140)                     & 297.69(201)                     & -2.61 & 0.0091 \\
Ryan Kelly       & 115.30(20)                      & 365.57(23)                     & -2.53 & 0.0113 \\
Paul Millsap     & 180.7(212)                      & 233.06(214)                     & -2.49 & 0.0129 \\
James Jones      & 213.43(28)                     & 482.53(38)                     & -2.41 & 0.0158 \\
Emmanuel Mudiay  & 156.89(95)                     & 224.53(179)                     & -2.37 & 0.0176 \\
Darrell Arthur   & 246.76(76)                     & 394.48(94)                     & -2.32 & 0.0202 \\
CJ Miles         & 186.81(113)                     & 259.75(194)                     & -2.25 & 0.0246 \\
Klay Thompson    & 144.16(257)                     & 179.24(309)                     & -2.21 & 0.0273 \\
Ersan Ilyasova   & 214.65(127)                     & 290.19(179)                     & -2.18 & 0.0292 \\
Andrew Wiggins   & 155.06(244)                     & 189.17(303)                     & -2.09 & 0.0367 \\
Sasha Vujacic    & 177.33(18)                     & 320.89(47)                     & -2.03 & 0.0427 \\
Al-Farouq Aminu  & 212.10(129)                      & 279.88(184)                     & -2.00    & 0.0457 \\
Brandon Knight   & 142.19(229)                     & 167.12(332)                     & -2.00    & 0.0458 \\
Bismack Biyombo  & 617.16(45)                     & 395.44(39)                     & 2.01  & 0.0447 \\
Tony Allen       & 435.00(63)                        & 296.14(84)                     & 2.07  & 0.0382 \\
Jarrett Jack     & 325.45(76)                     & 236.49(130)                     & 2.17  & 0.0299 \\
Cameron Payne    & 385.38(37)                     & 191.09(57)                     & 2.18  & 0.0291 \\
John Wall        & 194.63(205)                     & 152.61(290)                     & 2.32  & 0.0202 \\
Patrick Beverley & 348.02(83)                     & 221.66(103)                     & 2.40   & 0.0163 \\
JaVale McGee     & 383.65(46)                     & 164.92(25)                     & 2.48  & 0.0131 \\
Willie Reed      & 676.29(21)                     & 157.92(12)                     & 2.86  & 0.0042 \\
Jeff Withey      & 618.25(36)                     & 247.17(35)                     & 3.56  & 0.0004 \\
\hline
\end{tabular}
\caption{ (\ref{shotfrequency}) The entire NBA API dataset was used for the following analysis (Number of Players = 341). The second and third columns show the mean time difference (in "basketball seconds") between a player's shots after a miss and after a hit, respectively, with the number of attempts in parenthesis. The fourth and fifth columns show Z-scores found comparing these two means and the associated p-values using the hypothesis test $H_0:  \mu_{hit} = \mu_{miss}$ vs. $H_1: \mu_{hit} \neq \mu_{miss}$. The table above displays those players with significant p-values ($\alpha = 0.05$).}
\label{table:X4}
\end{table}

\begin{table}[h!]
\centering
\textbf{Mean Number of Dribbles Taken Before Next Shot After a Make vs. After a Miss}\par\medskip
\begin{tabular}{rrrrr}
\hline
Player Name        & Avg. \# of Dribbles After Make & Average \# of Dribbles After Miss & Z     & P-value  \\
\hline
James Ennis        & 0.57                     & 1.24                     & -2.85 & 0.0043 \\
Chase Budinger     & 0.25                     & 0.60                      & -2.53 & 0.0115 \\
Dante Cunningham   & 0.07                     & 0.36                     & -2.50  & 0.0123 \\
Anthony Bennett    & 0.28                     & 0.50                      & -2.17 & 0.0304 \\
Jeremy Lamb        & 1.58                     & 2.46                     & -2.08 & 0.0376 \\
Alan Anderson      & 0.98                     & 1.50                      & -1.98 & 0.0477 \\
Kyle Lowry         & 5.63                     & 4.81                     & 1.97  & 0.0485 \\
Kelly Olynyk       & 0.78                     & 0.52                     & 2.04  & 0.0418 \\
Kris Humphries     & 0.38                     & 0.23                     & 2.04  & 0.0414 \\
Ty Lawson          & 5.75                     & 4.95                     & 2.07  & 0.0388 \\
Ben Gordon         & 3.27                     & 2.37                     & 2.07  & 0.0385 \\
Chris Kaman        & 0.96                     & 0.68                     & 2.10   & 0.0357 \\
Aron Baynes        & 0.25                     & 0.10                      & 2.11  & 0.0352 \\
Tim Duncan         & 0.92                     & 0.68                     & 2.15  & 0.0315 \\
Trevor Booker      & 0.82                     & 0.52                     & 2.16  & 0.0308 \\
Chris Paul         & 6.52                     & 5.67                     & 2.21  & 0.0270  \\
Kyrie Irving       & 5.58                     & 4.84                     & 2.23  & 0.0260  \\
Tyreke Evans       & 5.11                     & 4.33                     & 2.23  & 0.0259 \\
Jonas Valanciunas  & 0.76                     & 0.52                     & 2.29  & 0.0222 \\
Jarrett Jack       & 6.08                     & 5.09                     & 2.29  & 0.0217 \\
Kemba Walker       & 5.69                     & 4.73                     & 2.34  & 0.0195 \\
Mo Williams        & 5.99                     & 4.81                     & 2.34  & 0.0195 \\
Donatas Motiejunas & 0.88                     & 0.62                     & 2.36  & 0.0185 \\
John Wall          & 6.22                     & 5.34                     & 2.37  & 0.0178 \\
Ryan Anderson      & 0.89                     & 0.66                     & 2.39  & 0.0169 \\
Marcin Gortat      & 0.58                     & 0.35                     & 2.39  & 0.0168 \\
Matt Barnes        & 0.68                     & 0.41                     & 2.48  & 0.0133 \\
Eric Bledsoe       & 5.16                     & 4.20                      & 2.52  & 0.0119 \\
Tony Parker        & 6.36                     & 5.16                     & 2.54  & 0.0110  \\
Lance Stephenson   & 3.51                     & 2.64                     & 2.65  & 0.0080  \\
Lebron James       & 5.25                     & 4.30                      & 2.66  & 0.0079 \\
Tobias Harris      & 1.63                     & 1.18                     & 2.68  & 0.0074 \\
Isaiah Thomas      & 4.69                     & 3.44                     & 2.88  & 0.0039 \\
Dwayne Wade        & 3.90                      & 3.06                     & 2.89  & 0.0038 \\
Caron Butler       & 0.99                     & 0.55                     & 3.00     & 0.0027 \\
Greg Monroe        & 1.37                     & 1.04                     & 3.00     & 0.0027 \\
Jeff Teague        & 6.55                     & 5.24                     & 3.11  & 0.0018 \\
Boris Diaw         & 1.97                     & 1.27                     & 3.16  & 0.0016 \\
Trey Burke         & 3.88                     & 2.70                      & 3.36  & 0.0008   \\
Russell Westbrook  & 5.86                     & 4.53                     & 4.08  & 0.0000     \\
\hline
\end{tabular}
\caption{(\ref{dribbles}) The NBA API dataset was filtered so each player tested had at least 15 misses and 15 hits (Number of Players = 281). The second and third columns show the mean number of dribbles a player takes before their next shot after a miss versus after a hit. The fourth and fifth columns show the  Z-scores found comparing these two means and the associated p-values using they hypothesis test $H_0:  \mu_{hit} = \mu_{miss}$ vs. $H_1: \mu_{hit} \neq \mu_{miss}$. The table above displays those players with significant p-values ($\alpha = 0.05$).}
\label{table:X5}
\end{table}

\begin{table}[h!]
\centering
\textbf{Mean Closest Defender Distance After a Make vs. After a Miss  }\par\medskip
\begin{tabular}{rrrrr}
\hline
Player Name        & Avg. Distance After Make & Avg. Distance After Miss & Z     & P-value  \\
\hline
Joe Harris         & 4.37(27)                          & 6.47(39)                          & -3.04 & 0.0024 \\
Jason Terry        & 4.46(120)                        & 5.31(162)                          & -2.80  & 0.0051 \\
Arron Afflalo      & 3.70(284)                           & 4.14(346)                          & -2.49 & 0.0127 \\
Nicolas Batum      & 5.31(156)                          & 6.13(240)                          & -2.28 & 0.0229 \\
Boris Diaw         & 4.40(170)                           & 5.21(208)                          & -2.25 & 0.0247 \\
Robert Sacre       & 3.33(83)                          & 4.05(108)                          & -2.11 & 0.0351 \\
Brian Roberts      & 4.30(124)                           & 4.87(189)                         & -2.03 & 0.0422 \\
Jared Dudley       & 6.04(148)                          & 5.36(153)                          & 2.01  & 0.0447 \\
James Ennis        & 5.24(47)                          & 4.05(70)                          & 2.05  & 0.0406 \\
Udonis Haslem      & 4.25(43)                          & 3.11(54)                          & 2.05  & 0.0401 \\
Nerles Noel        & 3.49(169)                          & 3.07(218)                          & 2.10   & 0.0356 \\
Chase Budinger     & 4.79 (60)                         & 3.87(95)                          & 2.12  & 0.0337 \\
Blake Griffin      & 4.85(421)                          & 4.36(421)                          & 2.17  & 0.0301 \\
Andrew Wiggins     & 3.67(327)                          & 3.25(411)                          & 2.17  & 0.0298 \\
Chandler Parsons   & 4.57(283)                          & 4.03(337)                          & 2.27  & 0.0234 \\
Brook Lopez        & 4.03(291)                          & 3.51(291)                          & 2.27  & 0.0230  \\
Tony Allen         & 4.03(151)                          & 3.07(156)                          & 2.33  & 0.0198 \\
Brandon Bass       & 3.81(196)                          & 3.32(223)                          & 2.39  & 0.0168 \\
Anthony Bennett    & 4.97(103)                          & 4.03(122)                          & 2.46  & 0.0140  \\
Carlos Boozer      & 3.64(292)                          & 3.22(270)                          & 2.49  & 0.0127 \\
Derrick Williams   & 4.57(116)                          & 3.76(124)                          & 2.53  & 0.0114 \\
Taj Gibson         & 2.92(185)                          & 2.43(184)                          & 2.60   & 0.0092 \\
Zach Randolph      & 3.10(305)                           & 2.67(316)                          & 2.72  & 0.0065 \\
Tyson Chandler     & 3.42(186)                          & 2.72(97)                          & 2.73  & 0.0064 \\
Jeremy Lamb        & 5.03(77)                          & 3.78(110)                          & 2.90   & 0.0038 \\
Charlie Villanueva & 5.77(87)                          & 4.68(112)                          & 3.00     & 0.0027 \\
Pau Gasol          & 4.08(369)                          & 3.49(392)                          & 3.09  & 0.0020  \\
Lamarcus Aldridge  & 4.27(454)                          & 3.80(544)                           & 3.10   & 0.0019 \\
Evan Turner        & 4.33(180)                          & 3.64(260)                          & 3.32  & 0.0009   \\
Derrick Favors     & 3.30(335)                           & 2.75(286)                          & 3.42  & 0.0006   \\
Kawhi Leonard      & 4.84(211)                          & 3.87(263)                          & 4.01  & 0.0001   \\
Marc Gasol         & 4.55(377)                          & 3.72(377)                          & 4.69  & 0.0000      \\
Quincy Acy         & 5.66(86)                          & 3.79(101)                          & 4.74  & 0.0000      \\
Marreese Speights  & 4.66(231)                          & 3.48(213)                          & 4.94  & 0.0000     \\
\hline
\end{tabular}
\caption{(\ref{defenderdistance}) The NBA API dataset was filtered so each player tested had at least 15 misses and 15 hits (Number of Players = 281).  The second and third columns show the mean distance (in feet) from the closest defender on their next shot after a miss versus after a hit, respectively. The fourth and fifth columns show the test statistic and its corresponding p-value for the test $H_0: D_{\text{hit}} - D_{\text{miss}} = 0$ versus $H_1: D_{\text{hit}} - D_{\text{miss}} \neq 0$, where $D$ is the mean closest defender distance. The table above displays those players with significant p-values ($\alpha = 0.05$). } 
\label{table:X6}
\end{table}

\begin{table}[h!]
\centering
\textbf{Shooting Performance Before Halftime versus Shooting Performance After Halftime }\par\medskip
\begin{tabular}{rrrrrrr}
\hline
Pre-Half Shots & Post-Half Shots & Shots/Half (Players) & $FGP_{pre}$ & $FGP_{post}$ & r   & P-value \\
\hline 
Last 3 & First 3 & 12486 (331)                                  & 0.4640       & 0.4646        & -0.0067     & 0.3322  \\
Last 4 & First 4 & 6992 (236)                                   & 0.4723       & 0.4715        & -5.22 $\times 10^{-5}$ & 0.4991  \\
Last 5 & First 5   & 3020 (129)                                   & 0.4682       & 0.4645        & -0.0682     & 0.04707 \\
Last 6 & First 6   & 1032 (78)                                   & 0.4767       & 0.4922        & 0.04969    & 0.7413 \\
All in Q2  & All in Q3   & 17585, 18886 (331)                           & 0.4672       & 0.4647        & 0.0108      & 0.7567 \\
\hline
\end{tabular}
\caption{(\ref{gamebreaks}) The NBA API dataset was filtered to include only data from quarters where players attempted at least three shots in quarters two and three before conducting each analysis. Columns 1 and 2 show the subset of shots from each half analyzed for each player in games where the player attempted a sufficient number of shots. Column three shows the total number of shots considered for each half (these values were the same for each half except in the last row, where there were 17,585 shots from before halftime, and 18,886 from after halftime). Columns four and five show the global field goal percentages from the subsets of shots pre-halftime and post-halftime, respectively. Column six shows the correlation between a player's individual pre-half field goal percentage and their corresponding post-half field goal percentage, and column six reports the accompanying p-value for the test ($H_0: r = 0$ vs $H_1: r < 0$)} 
\label{table:X7}
\end{table}

\clearpage

\section{Figure Captions}
\begin{figure}[h!]
    \centering
    \caption{(\ref{analysisofrunssims}) The output of an FDR \citep{benjamini_controlling_1995} analysis for values of $\delta$ ranging from 0 to 0.6. For 21 different values of $\delta$, 10 simulations were generated, for a total of 210 simulations. The Benjamini-Hochberg output (for $\alpha = 0.05$) comes from analyzing the p-values of the Wald-Wolfowitz test of each player's simulated shot results (Number of Players = 438), with the simulations generated using equation (2). A significant number of discoveries (more than 22) is first seen at at $\delta = 0.39$. The red line marks the threshold for a significant number of declared discoveries.} 
    \label{fig:my_label}
\end{figure}

\begin{figure}[h!]
    \centering
    \caption{(\ref{globaltestsim}) The global test statistics (using equation (1)) from each of the 210 simulations also used in Section 3.6.1 (Number of Players = 438). This analysis was used in Section 3.2 to globally test if $p_h - p_m > 0$. As $\delta$ increased, the global test statistic increased, gradually growing more significant. Clear evidence of a global hot hand $(T > 0)$ is seen around $\delta = 0.18$ and beyond.}
    \label{fig:my_label2}
\end{figure}

\clearpage
\section{Figures} 

\begin{figure}[h!]
    \centering
    \textbf{Fig. 1: Simulation of Wald-Wolfowitz Test}\par\medskip
    \includegraphics[scale = 0.8]{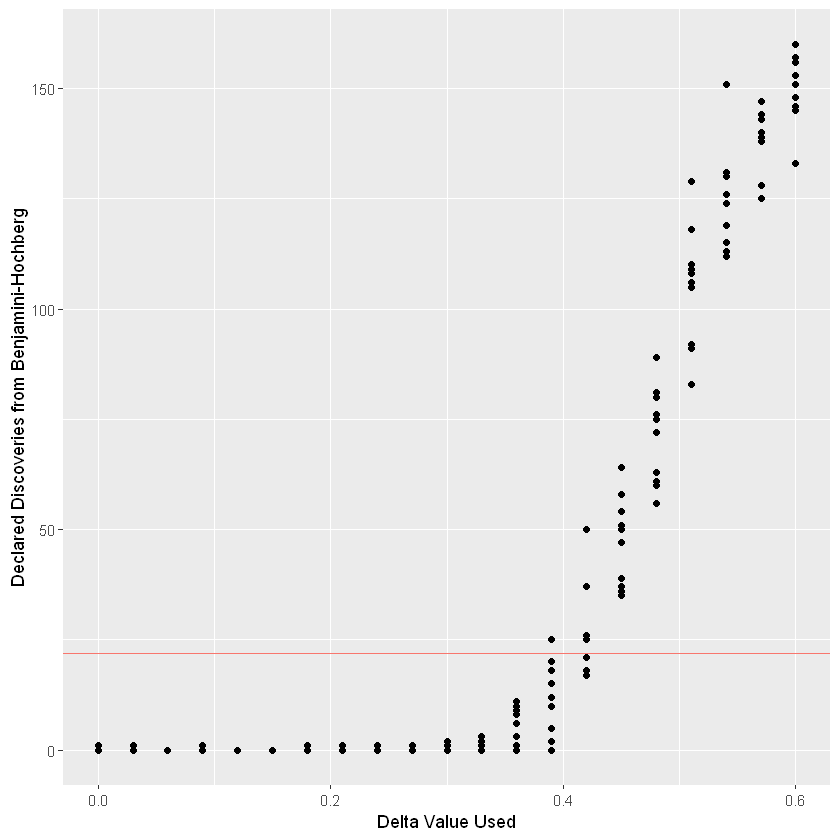}
    
\end{figure}

\begin{figure}[h!]
    \centering
    \textbf{Fig. 2: Simulation of Global t-test}\par\medskip
    \includegraphics[scale = 0.8]{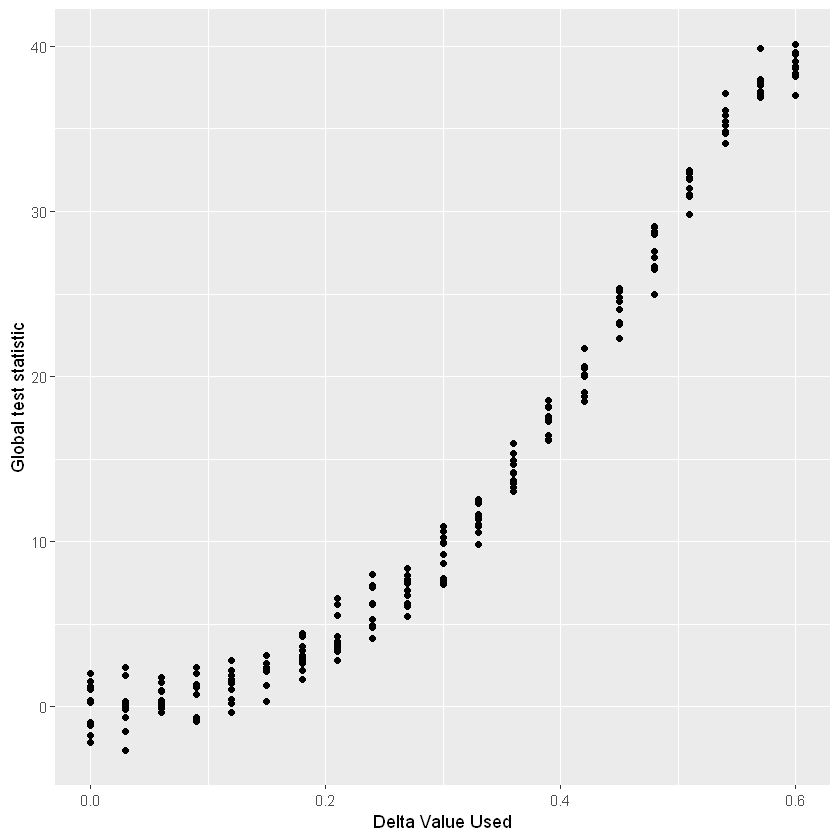}
    
\end{figure}

\end{document}